\documentclass[preprint,showpacs]{revtex4}
\usepackage{graphicx}
\usepackage{bm}
\usepackage{color}
\usepackage{amsmath}
\usepackage{natbib}
\usepackage{latexsym}

\newcommand{\vect}{\boldsymbol}

\begin{document}

\title{Hot Fluids and Nonlinear Quantum Mechanics}

\author{Swadesh M. Mahajan}
\email{mahajan@mail.utexas.edu}
\author{Felipe A. Asenjo}
\email{faz@physics.utexas.edu}
\affiliation{Institute for Fusion Studies, The University of Texas at Austin, Texas 78712, USA.}

\date{\today}

\begin{abstract}
It is shown that a hot relativistic fluid could be viewed as a collection of self-interacting quantum objects. They
obey a nonlinear equation which is a modification  of the quantum equation
obeyed by elementary constituents of the fluid. A uniform phenomenological prescription, to affect the  quantum transition from a corresponding classical system, is invoked to  derive the nonlinear Schr\"odinger, Klein--Gordon, and Pauli--Schr\"odinger equations. It is expected that the
created nonlinear quantum mechanics would advance, in a fundamental way, both the conceptual understanding and
computational abilities, particularly, in the field of extremely high energy-density physics.
\end{abstract}


\maketitle

\section{Introduction}

This paper is an attempt to construct a nonlinear quantum mechanics (NQM) that will adequately capture the thermal interactions of
an assembly of elementary particles.  Using classical nomenclature,  such an assembly may be called a fluid (it can have charge and other internal degrees of freedom). The  object  that emerges on ``quantization" of this fluid will be, then,  given the name ``{\it Fluidon}". This paper, thus, builds a quantum theory of ``Fluidons".

It is helpful to provide some underlying  historical background  for this work. Exploring the  classical-quantal interplay to create, perhaps, a basis  for a classical style visualization of quantum mechanics has attracted a variety of researchers ever since the dawn of quantum mechanics.  Although, to the best of our knowledge, this quest has not yielded very fruitful insights, the interest  in this deep and fascinating
subject has waxed and waned but it has never disappeared. Two main trends are easily discernible in the literatue:

1) {\it Quantum to Classical}: The first attemps were focused in the construction of a fluid-like system from the standard equations of quantum mechanics, the Schr\"odinger, the Pauli--Schr\"odinger, the Klein--Gordon, the Feynman-Gellmann and the Dirac equations \cite{madelung, bohm, takabayasi,cufaro,aron,lin}. In the earlier  years of quantum mechanics, this enterprise had a limited scope;  the investigators were content to devise appropriate fluid-like variables obeying the ``expected'' fluid-like dynamics epitomized in  the continuity and  the  force balance equations . Quantum mechanics entered  the latter through a  variety of  ``quantum forces'' proportional to powers of $\hbar$; the simplest easily recognizable quantum force came  from the well-known Bohm potential. The fluidized system, of course, was equivalent to the original quantum one. In more recent years, however, an impressive regeneration of the fluidization project was fueled by  attempts to investigate  the  collective macroscopic motions accessible to a fluid (plasma) whose elementary constituents follow the laws of quantum rather than classical dynamics. Much progress has been made in the direction of building and  exploring macroscopic theory  of quantum plasmas \cite{Fro,haas,marklund1,marklund2,vortical,asenjorqp} as opposed to the earlier efforts that mostly consisted of casting quantum mechanics into a fluid-like mould. New plasma phenomena, originating in the quantum nature of the constituent particles, have come into light

2) {\it Classical to Quantum}: The reverse path has also been heavily travelled-  there exist many "derivations"  of the elementary equations of quantum mechanics from classical kinetic (phase space) and fluid (configuration space) theories \cite{kania,pesci1,pesci2,pesci3}; the quantum equations  follow when very specific choices of the  distribution function (or the fluid variables) are made. In most of these investigations, barring a very recent paper \cite{carbonaro}, the  sought after result was the standard linear equations of quantum mechanics, the Schrodinger \cite{kania,pesci1}, the Pauli--Schr\"odinger \cite{pesci2}, the Klein--Gordon and the Feynman--Gellmann version of the Dirac equation \cite{pesci3}.

This paper would fall in the second category.
It is somewhat naive  to believe  that the full expression of the state of a fluid with collective features like temperature (and an internal energy) could be captured in  an elementary linear quantum equation. We will, therefore, seek a quantum description
appropriated to a hot relativistically fluid. Not unexpectedly, this description turns out to define a nonlinear quantum mechanics; the nonlinearity is an expression of the internal energy of the fluid. The latter insures that the fluid is not just a collection of  noninteracting ``quantum" particles.

Our procedure is  fully inspired by the very original  formulation of quantum mechanics where, for example, the Schr\"odinger  and the Klein--Gordon equations were derived by the respective identifications
\begin{equation}
E_{nonrel} = i\hbar\partial_t,   \,\,\, \mathbf{p}= \frac{\hbar}{i}\nabla
\end{equation}
\begin{equation}
p^\mu = \frac{\hbar}{i}\partial^\mu
\end{equation}
where $p^\mu=[E, \mathbf{p}]$ is the momentum four-vector, $E_{nonrel} = E-m c^2$. The Schr\"odinger equation can be viewed as a phenomenological equation obtained by the above mentioned identification.

Our objective is to seek a similar, hopefully unique, quantum prescription to develop a (phenomenological) quantum theory of a hot fluid. We will illustrate the general development by working with the simplest example of such a system - a perfect isotropic field.

\section{Hot Relativistic Perfect  Fluid}

The energy momentum tensor for an isotropic  relativistic (both thermally and kinematically) perfect fluid composed of particles with no  internal degree of freedom may be written as
\begin{equation}\label{EnMom}
T^{\mu\nu}=\frac{n}{w} f\,p^\mu p^\nu+\Pi\, \eta^{\mu\nu}\, ,
\end{equation}
where $p^\mu$ is the fluid kinematic momentum, $\Pi$ is the pressure, $f$ is the enthalpy, $n$ is the invariant number density, and $w$ is a constant with dimensions of mass to insure that $T^{\mu\nu}$ has the dimensions of energy density. We choose the Minkowsky signature tensor to be $\eta^{\mu\nu}=[-1,1,1,1]$  \cite{weinbook}. Notice that $T^{\mu\nu}$ in Eq.~\eqref{EnMom} is the same as the expression in standard textbooks  because  $n f w = U+\Pi$, where $U=T^{00}$ is the proper energy density.

The invariant density $n$ appears in the continuity equation obeyed by the fluid
\begin{equation}\label{continuity}
  \partial_\mu(np^\mu)=0\, .
 \end{equation}
For a fluid that has a relativistic temperature in addition to a relativistic directed motion, the bare kinematic momentum
ceases to be the right dynamic variable in terms of which the equations of motion have a ``simple structure"; it also ceases to be the appropriate variable in terms of which, via the minimal prescription $p^\mu\rightarrow  p^\mu-eA^\mu/c$, one can  correctly incorporate electromagnetism into the fluid dynamics ($A^\mu$ is the vector potential). It was shown in a 2003 paper \cite{mah2003} that,
for a relativistically hot fluid, the thermally modified ``effective momentum"
 \begin{equation}\label{P}
 {p^\mu}_{eff}\equiv P^\mu=f\, p^\mu\, ,
\end{equation}
is endowed with both of these desirable properties. The effective momentum, obviously, reflects the thermal enhancement of the rest mass by a factor $f$ (that depends only on the thermal distribution of the particles). Using \eqref{P}, one may write the Eqs.~\eqref{EnMom} and \eqref{continuity} as
\begin{equation}\label{energyM2}
T^{\mu\nu}=g\, P^\mu P^\nu+\Pi\, \eta^{\mu\nu}\, ,
\end{equation}
\begin{equation}\label{P2}
\partial_\mu(g P^\mu)=0\, .
\end{equation}
where $g=n/fw$.

\subsection{Quantization for the effective momentum}
It is proposed that the transition to the equivalent ``quantum mechanics of a Fluidon" is affected by the prescription
\begin{equation}\label{quantprescrip}
 g P^\mu P^\nu \Longrightarrow g\left(P^\mu P^\nu+\frac{\hbar}{2i}\partial^\mu \frac{\hbar}{2i}\partial^\nu\ln g\right).
\end{equation}
This prescription looks both similar to, and distinct from the conventional prescription $p^\mu=-i\hbar \partial^\mu$ invoked in particle quantum mechanics. Some comments on this topic will be offered after the required equations are obtained. The transformed  energy-momentum tensor [via prescription \eqref{quantprescrip}]
\begin{equation}\label{energyM3}
T^{\mu\nu}_q=g\, P^\mu P^\nu+\Pi\, \eta^{\mu\nu}-\frac{\hbar^2}{4}g\partial^\mu\partial^\nu\ln g\, ,
\end{equation}
has been labelled with the subscript $q$ [to differentiate it from \eqref{energyM2}]. We will first work out the quantum dynamics of a free Fluidon, i.e, when there are no external forces ; the only forces that the Fluidon experiences are inertial and thermal. Then, the equation of motion
\begin{equation}\label{eqmotionpre}
0= \partial_\mu T^{\mu\nu}_q,
 \end{equation}
in conjunction with  the continuity equation~\eqref{P2}, yields $(\zeta=\ln ({n/f}))$
\begin{equation}\label{eqmotion}
0= \frac{1}{g}\partial^\nu\Pi+ P^\mu\partial_\mu P^\nu-\frac{\hbar^2}{4}\partial^\nu\left(\frac{1}{2}\partial_\mu\zeta\partial^\mu\zeta+\Box\zeta\right)
  \end{equation}
Notice that  the  quantum prescription \eqref{quantprescrip} was so chosen that the term proportional to $\hbar$ appears as a perfect four-gradient in the  equation of motion. The other two terms on the right hand side of ~\eqref{eqmotion}
will also become perfect gradients provided that the effective momentum is  an irrotational field (vorticity free),
\begin{equation}\label{MomtoAction}
  P^\mu=\partial^\mu S,
\end{equation}
and there exists an equation of state of the form
\begin{equation}\label{eqstate}
  \Pi=\Pi(g)
\end{equation}
When both these conditions are satisfied, the equation of motion  may be  integrated to yield
\begin{equation}\label{eqquant1}
 d=\bar\Pi+\frac{1}{2}(\partial_\mu S)(\partial^\mu S)-\frac{\hbar^2}{4}\left(\frac{1}{2}\partial_\mu\zeta\partial^\mu\zeta+\Box\zeta\right)
\end{equation}
where $d$ is an integration constant and $\bar\Pi(g)$ is determined by solving $g d\bar\Pi/dg=d\Pi/dg$. In terms of $S$ and $\zeta$, the continuity equation reads
\begin{equation}\label{P3}
  0=\partial_\mu\partial^\mu S+\partial_\mu\zeta\partial^\mu S\, ,
\end{equation}
In two direct and straightforward algebraic manipulations, consisting of first introducing an intermediate variable,
\begin{equation}
\Omega=\frac{\zeta}{2} +\frac{i}{\hbar}S\, ,
\end{equation}
and then defining the wave function
\begin{equation}\label{Madelung}
\ln\Psi=\Omega \Longrightarrow \Psi=\sqrt{\frac{n}{f}}\,\, e^{iS/\hbar},
\end{equation}
we derive form \eqref{eqquant1} and \eqref{P3}, the sought after quantum equation
\begin{equation}\label{eqquant2}
\left[{\hbar}^2 \partial_\mu\partial^\mu-\bar\Pi+d\right]\Psi=0,
\end{equation}
obeyed by the Fluidon. Interestingly, the probability density
\begin{equation}\label{probdensity}
\Psi^*\Psi=\frac{n}{f}
\end{equation}
associated with the Fluidon is not the original number density of the fluid elements; it is modulated by $f$, representing the thermal content of the fluid.

\subsection{Interpretation-Conjectures}

The Fluidon quantum equation is likely to open exciting new channels for exploring many body physics via a relatively simple nonlinear equation. We will attempt to extricate some obvious features of the  rich ``content" of  the quantum equation~\eqref{eqquant2}:

1) If the fluid began  with no internal thermal energy (zero temperature), its pressure  $\Pi$ (hence $\bar\Pi$) vanishes and the enthalpy factor $f\rightarrow1$. The nonlinear  quantum equation, then, reduces  to the standard linear Klein--Gordon (KG) equation
\begin{equation}\label{eqquant3}
[{\hbar}^2 \partial_\mu\partial^\mu-m^2]\Psi=0,
\end{equation}
when the constant $d$ is chosen to be $-m^2$. The wave function$ \Psi$ also simplifies to
\begin{equation}\label{Madelung}
\Psi=e^{\Omega}=\sqrt{n}\,\,  e^{iS/\hbar}
\end{equation}
so that $n=\Psi^*\Psi$ represents the  conventional probability density. Thus the  quanta, that emerge when the quantization prescription~\eqref{quantprescrip} is applied to a fluid with no internal energy (and with a vorticity free kinematic momentum), obey the linear KG equation, exactly like a free  elementary particle with the relativistic energy-momentum relation $E^2=p^2+m^2$. Thus the zero temperature fluid may be perceived as consisting of non-interacting free quanta. This identification/recognition defines the baseline for further discussion.

2) An exciting new vista opens up when we switch on the internal thermal energy;  that will reinstate both $\bar\Pi(g)$ and $f$. The first direct consequence is that $g=(1/w)n/f= \Psi^*\Psi/w$ implying $\bar\Pi(g)\equiv \bar\Pi(\Psi^*\Psi)$. The dynamics associated with the Fluidon (born out of a hot fluid) is, then, governed by the nonlinear KG equation
\begin{equation}\label{eqquantnonl}
\left[{\hbar}^2 \partial_\mu\partial^\mu-\bar\Pi(\Psi^*\Psi/w)+d\right]\Psi=0,
\end{equation}
where the nonlinear term is displayed as a function of the probability density $\Psi^*\Psi$. The exact expression for the nonlinear term will, of course, depend on the equation of state. {\it As remarked earlier, the wavefunction $\Psi$ no longer measures the conventional probability density $n$, but a thermally weighted effective density}.

3) The transition from a cold to a hot fluid translates into a profound transition for the respective fluid quanta - the cold Fluidons are non-interacting and follow a linear  KG equation, while the hot Fluidons are interactive and obey a KG equation with an added nonlinear term;  the nonlinear term reflects the internal thermal energy. One could, perhaps, dare to venture that nonlinear quantum mechanics  constitutes a proper ``replacement''  of the standard linear quantum mechanics  for  describing an ``elementary" constituent of  a system of  thermally interacting particles.

\subsection {The Nonlinear Term }
To get a feel for the nonlinear term, let us take a specific equation of state
\begin{equation}\label{eqstate}
\Pi= a g^\Gamma\longrightarrow \bar\Pi= \frac{a\Gamma}{\Gamma-1}g^{\Gamma-1}
\end{equation}
This choice converts Eq.\eqref{eqquantnonl} into
\begin{equation}\label{eqquantnonlex}
\left[{\hbar}^2 \partial_\mu\partial^\mu- \lambda(\Psi^*\Psi)^{\Gamma-1}+d\right]\Psi=0,
\end{equation}
where $\lambda= (a/ w^{\Gamma-1}) \Gamma/({\Gamma-1})$ is a fluid specific constant.  For normal matter, $\Gamma>1$ implying that both $\lambda$, and  the exponent of the nonlinear term  are positive definite. For exotic matter, $\lambda$ need not be positive definite. Investigation of this nonlinear equation is likely to bring out very interesting physics. For instance, for  $\Gamma=2$, Eq.~\eqref{eqquantnonlex}  corresponds  to the highly investigated $\Psi^4$ theories that have been invoked as models for spontaneous symmetry breaking (when the vacuum does  not have the symmetries of the Lagrangian). The choice $d= - m^2$,  leads to a nonlinear extension of the KG field, but if $d=\mu^2>0$,  $\lambda>0$, the field develops a finite vacuum expectation value.

It is interesting to note that the Fluidon energy momentum relation is fundamentally different from that of a linear K-G particle. It is explicitly seen  for a particle with real mass $(d=-m^2)$, by  substituting the energy- momentum eigenfunction
\begin{equation}\label{Momeigenfunction}
\Psi=\Psi_0   e^{i \mathcal{\mathcal{P}}_\mu x^\mu/\hbar},
\end{equation}
where $\mathcal{P}_\mu=[\mathcal{E}, \bf{P}] $ is the four momentum and $\Psi_0$  is a constant amplitude, into Eq.\eqref{eqquantnonl}. We find
\begin{equation}\label{NonlinEnMom}
\mathcal{E}^2=\mathcal{\bf P}^2+m^2 + \lambda {\Psi_0}^{2(\Gamma-1)};
\end{equation}
the last amplitude dependent term is the expression of nonlinearity. This contribution could be viewed as an effective mass and could dominate the intrinsic rest mass for sufficiently high pressure.

In the context of preceding discussion, Fluidons could be viewed as ``many body" quanta that carry (through the nonlinearity in the equation) the extra information characteristic of an interacting many body system like a hot fluid. The Fluidons, governed by the nonlinear KG equation \eqref{eqquantnonlex}, are thermally  interacting  spin-less bosons. In the next section we will talk about Fluidons that emerge when the hot  fluid has additional internal structure.

\subsection{Switching on the Electromagnetic Field}

If the hot fluid is  composed of electrically charged Fuidons, then the  electromagnetic interaction can be readily worked in. The most obvious and straightforward procedure is to  change the equation of motion $ \partial_\mu T^{\mu\nu}_q=0$ to
\begin{equation}\label{eqmoem1}
\partial_\mu T^{\mu\nu}= q\frac{n}{w}F^{\mu\nu}p_\mu= q g F^{\mu\nu}P_\mu
\end{equation}
where $q$ is the elementary charge, $F^{\mu\nu} = \partial^\mu A^\nu- \partial^\nu A^\mu$ is the Faraday tensor, and $A^\mu=[\phi, \mathbf A]$ is the electromagnetic four potential. Following exactly the procedure laid out in subsection IIB - but demanding that, now, the combination $P^\mu+qA^\mu$ be a  perfect gradient
\begin{equation}\label{MomtoActionem}
  P^\mu+ q A^\mu=\partial^\mu S,
\end{equation}
one ends up deriving the equation of motion of the KG Fluidon subjected to an electromagnetic field,
\begin{equation}\label{eqmoem2}
\left[-(i\hbar\partial_\mu-qA_\mu)(i\hbar\partial^{_\mu}-qA^\mu) - \lambda  \bar\Pi(\Psi^*\Psi) +d\right]\Psi=0,
\end{equation}
where we have switched back to the general form of the nonlinear term $\lambda  \bar\Pi(\Psi^*\Psi)$.
Indeed, we would have gotten this result without any extra calculation - by simply invoking the minimal coupling prescription (pertinent to a hot relativistic fluid) that was worked out in Ref.~\cite{mah2003}, $ gp^\mu=P^\mu \rightarrow P^\mu+q A^\mu$, precisely what transpired in going from \eqref{MomtoAction} to \eqref{MomtoActionem}. When this replacement is made, the effective fluid momentum $P^\mu$ will, in general,  have vorticity but it is the generalized momentum, $P^\mu+q A^\mu$, that is vorticity free \cite{mah2003}.

 \subsection{Non relativistic Limit- Landau Ginzburg Model}

Here we follow the opposite of  the standard practice of building the theory from the simpler non-relativisic to the more involved relativistic systems. The non relativistic version  of the nonlinear KG equation follows by recalling that, in this limit,  the rest mass is much greater than the kinetic and thermal energies ($\mathcal{E}\simeq m)$. Then
\begin{equation}\label{nonrelEn}
\mathcal{E}^2-m^2= (\mathcal{E}-m)(\mathcal{E}+m)\simeq 2m E_{nonrel}
\end{equation}
where $E_{nonrel}$ is the energy without the rest mass. Substituting \eqref{nonrelEn} into \eqref{NonlinEnMom}, switching on the electromagnetic field, and affecting $E_{nonrel} \rightarrow i\hbar \partial t, \mathcal{P}_{k}\rightarrow -i\hbar\partial_k$ (latin indices go from 1-3 as greek indices go from 0-3), we obtain
\begin{equation}\label{}
 i\hbar\frac{\partial\Psi}{\partial t}=\left[-\frac{\hbar^2}{2\mu}\left(\partial^{k}-\frac{iq}{\hbar}A^{k}\right)^2+q\phi+\lambda|\Psi^*\Psi|^{(\Gamma-1)}\right]\Psi,
\end{equation}
a nonlinear Schr\"odinger equation satisfied by a non relativistic Fluidon interacting with an electromagnetic field. For $\Gamma=2$, the Schrodinger  Fluidon follows  the same equation as the Super-electron of the Landau-Ginzburg model.

For arbitrary $\Gamma$, and even for equations of state  more complicated than the adiabatic law \eqref{eqstate}, the non relativistic Fluidon will always will be described by  what may be called a generalized Nonlinear Schr\"odinger equation (NSE) where the nonlinear term can be an arbitrary function $f(|\Psi^*\Psi|)$ \cite{mahnanber}. The NSE has been one of most investigated equations and has yielded results of immense significance spanning  classes of stable localized solutions (including solitons)\cite{mahnanber}. Thus the investigative apparatus to study Flulidon dynamics is already in place.
On the other hand, following the opposite path, several attempts have been  performed to fluidize nonlinear Schr\"odinger equations  \cite{spiegel,ercolani,nassar,non} finding some related results.

\section{Hot fluids with an internal degree of freedom}

The single component KG Fluidon represents a many-body manifestation of a system of thermally interacting spinless bosons. Being
relativistic counterpart  of the Schr\"odinger Fluidon, it may be invoked  to understand the  many-body behavior of an assembly of relativistic electrons, but such a description is bound to be inadequate since the model lacks, {\it inter  alia}, the intrinsic spin of the constituent particles.

In this section we will carry out fluid-quantization for a perfect fluid with intrinsic spin.
Ideally, in analogy with the preceding sections on the nonlinear KG,
one would expect to  be able to derive a nonlinear Dirac equation for the spin half particles.
Although the algebra for such an exercise is straightforward (though long and tedious),
there are some conceptual problems  that we have not yet  been able to resolve, Consequently  we will
hold  the presentation of the equation for a  relativistic spin half particle  for a later publication. Instead we will
derive a phenomenological nonrelativistic equation for a spin-half Fluidon; naturally we expect it to be a
nonlinear Pauli--Schr\"odinger equation.

\subsection{The non-relativistic  spin-1/2 Fluidon - Nonlinear Pauli--Schr\"odinger (PS) equation}

In the non-relativistic limit, the vector part of classical  energy-momentum conservation law, $0= \partial_\mu T^{\mu\nu}$, reduces to
\begin{equation}\label{eqPaemt}
  \partial_t T_{0k}+\partial_jT_{jk}=F_k\, ,
\end{equation}
where $T_{0k}=mn v_k$ are the mixed, while $T_{jk}=m n v_jv_k+p\delta_{jk}$ are the space components of the energy-momentum tensor (corresponding to a spinless partcile). In \eqref{eqPaemt}, $F_k$ represents all the external forces acting on the system, and $p$ is the fluid pressure that obeys an equation of state  $p=p(n)$. We have chosen to work with the velocity rather than the momentum variables for this section.

In order to introduce appropriate spin induced modifications to the energy momentum tensor, we have to draw from the
pioneering work of Takabayasi \cite{TakabayasiEMT}, and the procedures worked out in \cite{holland}. The first part we borrow is
the spin induced stress
\begin{equation}\label{spinstress}
\sigma_{jk}=(\hbar^2/4m)n\partial_j\Sigma_i\partial_k\Sigma_i\, ,
\end{equation}
where $\Sigma_i=\varphi^\dag\sigma_i\varphi$ is the spin field associated with the  Fluidon, $\sigma_i$  are the Pauli matrices, and $\varphi$ is the normalized Fluidon spinor ($\varphi^\dag\varphi=1$, $\Sigma_i \Sigma_i=1$).  This term  can be viewed as phenomenological and is  chosen to reproduce the linear PS equation when the fluid pressure is zero. One is, of course, seeking the equation that $\varphi$ obeys.

Adding $\sigma_{jk}$ to the classical stress and  applying the non relativistic limit of the  quantum prescription \eqref{quantprescrip}, we obtain, to the leading order, the appropriate quantum stress
\begin{eqnarray}
 T_{0k}^q&=& mn v_k\, ,\nonumber\\
  T_{jk}^q&=& mn v_jv_k-\frac{\hbar^2}{4m}n\partial_j\partial_k\ln{n}+p\delta_{jk}+\sigma_{jk}\, ,
\end{eqnarray}


With this input, the use of he continuity equation
\begin{equation}\label{conPA}
  \partial_{t}n+\partial_j(n v_j)=0\, ,
\end{equation}
and the identity $(1/n)\partial_j(n\partial_j\partial_k\ln n)=\partial_k(2\partial_j^2\sqrt n/\sqrt n)$,
the quantum translation of Eq.~\eqref{eqPaemt} becomes
\begin{equation}\label{eqPaemt2}
 m \partial_t v_{k}+mv_j\partial_j v_{k}=F_k+\frac{\hbar^2}{2m}\partial_k\left(\frac{\partial_j^2\sqrt n}{\sqrt n}\right)-\frac{\partial_j\sigma_{jk}}{n}-\frac{\partial_kp}{n}\, ,
\end{equation}
where $F_k=e E_k+e\varepsilon_{klm}v_l B_m+(2\mu/\hbar)\Sigma_l\partial_kB_l$ ($\mu=e\hbar/2m$),  is composed of the standard Lorentz force, and the additional force originating in the internal spin interacting with the magnetic field \cite{TakabayasiEMT}. The electromagnetic fields may, as usual, be expressed in terms of the scalar ($\phi$) and vector potentials ($A_i$):  $E_k=-\partial_k\phi-\partial_t A_k$, and $B_i=\varepsilon_{ijk}\partial_jA_k$.

For the fluid equations to be complete, one needs an equation for the evolution of the spin field $\Sigma$. Fortunately, such an equation was derived in Ref.~\cite{TakabayasiEMT}
\begin{equation}\label{spin}
  (\partial_t+v_j\partial_j)\Sigma_i=\frac{2\mu}{\hbar}\varepsilon_{ijk}\Sigma_jB_k+\frac{1}{m n}\varepsilon_{ijk}\Sigma_j\partial_l(n\, \partial_l \Sigma_k)\, .
\end{equation}

In order to arrive at the  quantum Fluidon equation, that is equivalent to the fluid system described by \eqref{eqPaemt}-\eqref{spin}, we first
definine the irrotational (vorticity-free)  combination ($S$ is a scalar to be later identified with action)
\begin{equation}\label{}
  \partial_j S=m v_j+i\hbar \varphi^\dag\partial_j\varphi+e A_i\, ,
\end{equation}
that converts  the  system of Eqs.~\eqref{eqPaemt2}-\eqref{spin} into $\partial_k \chi=0$ (see Ref.~\cite{holland}), where
\begin{equation}\label{grad0}
\chi=\partial_t S-i\hbar\varphi^\dag\partial_t\varphi+\frac{1}{2}m v_i v_i-\frac{\hbar^2\partial_j^2\sqrt n}{2m n}+\frac{\hbar^2}{8m}\partial_i S_j\partial_i S_j+\frac{2\mu}{\hbar}S_jB_j+e\phi+\bar\Pi\, ,
\end{equation}
and where the function $\bar\Pi=\bar\Pi(n)$, just as in the KG case, is related to the pressure via
\begin{equation}\label{}
n  \frac{\partial\bar\Pi}{\partial n}=\frac{\partial p}{\partial n}\, .
\end{equation}

Choosing the solution $\chi=0$, and literally repeating the KG procedure,  Eqs.~\eqref{conPA}, \eqref{spin} and \eqref{grad0} are  found to be equivalent to the non-linear PS equation
\begin{equation}\label{}
  i\hbar\frac{\partial\Psi}{\partial t}=\left[-\frac{\hbar^2}{2m}\left(\partial_k-\frac{ie}{\hbar} A_k\right)^2 +\mu B_j\sigma_j+e\phi+\Pi\left(\Psi^\dag\Psi\right)\right]\Psi\, ,
\end{equation}
for the wavefunction
\begin{equation}\label{}
  \Psi=\sqrt n e^{iS/\hbar}\varphi\, ,
\end{equation}
where $\varphi$ is a two component spinor; the norm of the wave function, $\Psi^\dag\Psi=n$, is the Fluidon density. For the polytropic equation of state, $p=n^\Gamma $, the nonlinearity takes the form $\Pi=\lambda n^\Gamma=\lambda (\Psi^\dag\Psi)^\Gamma$ where$\lambda=  a\Gamma/({\Gamma-1})$

\section{An exact solution of the Nonlinear Klein--Gordon (NKG) - Maxwell system}

The primary aim of this paper was  developing the  conceptual foundations  for deriving the equations of nonlinear quantum mechanics governing a variety of fluidon fields. This was mostly accomplished in Sec. II and Sec. III. Now we present an exact nonlinear solution that describes the propagation of a circularly polarized electromagnetic wave in a fully relativistic hot spinless charged fluid. The physics of this system is contained in the  NKG equation in the presence of an electromagnetic field [Eq.~\eqref{eqmoem2} with $d=-m^2$]
 \begin{equation}\label{eqmoem22}
[(i\hbar\partial_\mu-qA_\mu)(i\hbar\partial^{_\mu}-qA^\mu) + \lambda\bar\Pi(\Psi^*\Psi) +m^2]\Psi=0,
\end{equation}
and the Maxwell equation
\begin{equation}\label{maxwell2}
 \partial_\mu\partial^\mu A^{\nu} =-4\pi J^{\nu}\, ,
\end{equation}
where $J^{\nu}$ is the current associated with the KG fields satisfying \eqref{eqmoem22}. The  four-vector potential of a circularly polarized electromagnetic wave, propagating along the $z$ direction, it only has spatial transverse components. It may be represented as $(A_0=0=A_z)$
\begin{equation}\label{circ}
\vect A=A\left[\hat e_x \cos(kz-\omega t)-\hat e_y\sin(kz-\omega t)\right]\, ,
\end{equation}
where $\omega (k)$ is the frequency (wave number), and $A$ is the constant amplitude of the wave. For this wave it is easy to note that $\vect A\cdot\vect A= A^2$, $\nabla\cdot \vect A=0$, and the current  density to complete the Maxwell equation is
\begin{equation}
\vect J=-\frac{q^2 (\Psi^*\Psi) }{m}\vect A= -\frac{q^2 n}{mf}\vect A
\end{equation}
where where the normalization $\Psi^*\Psi=n/f$ \eqref{probdensity} has been invoked.

With these simplifications, and the further assumption that the  wave function has only time and no space dependence, i.e,
\begin{equation}\label{solLAKG}
  \Psi=\sqrt{\frac{n}{f}}\, \exp\left(\frac{i\gamma}{\hbar}t\right)\, ,
\end{equation}
the resulting NKG equation
\begin{equation}\label{LAKG2}
  \hbar^2\frac{\partial^2\Psi}{\partial t^2}+q^2 A_0^2 \Psi+\lambda\bar\Pi(\Psi^*\Psi)^\Psi+m^2\Psi=0\, ,
\end{equation}
and the relevant part of the Maxwell equation
\begin{equation}\label{Maxw}
  \frac{\partial^2 \vect A}{\partial t^2}-\nabla^2\vect A=4\pi\vect J\, ,
\end{equation}
are, simultaneously, solved provided the conditions ($\omega_p=\sqrt{4\pi q^2 n/m}$ is the plasma frequency)
\begin{equation}
\gamma^2=m^2+q^2A_0^2+\lambda\bar\Pi\left(\frac{n}{f}\right)\, ,
\end{equation}
\begin{equation}\label{dispLA}
  \omega^2-k^2=\frac{\omega_p^2}{f}\, ,
\end{equation}
are satisfied. The former is the effective fluidon energy ($E=\gamma mc^2$) appropriately  modified by thermodynamics; the modifications enter  the expression through the nonlinear term proportional to $\lambda$. The latter reflects a profound change in the dispersion relation (the standard one will be  $\omega^2-k^2= {\omega_p^2}$) brought about by the fact that for the fluidon the norm of the wave function stands for an effective density, that equals  the standard density divided by the enthalpy factor $f$. Since for normal  matter  $f >1$ (for highly relativistic matter $f\gg1$), the new dispersion relation implies a reduced  effective plasma frequency  revealing the possibility of a pressure induced transparency. A hot  fluidon for the same rest mass density $n$ will be more transparent to light than a cold system. For highly relativistic temperatures, the effect can be enormous.

\section{Summary}

A phenomenological quantum prescription, very similar to the standard quantum prescription ($p^\mu = -i \hbar\partial^\mu$), applied to a hot relativistic fluid yields a thermal quantum object (fluidon) that obeys a nonlinear quantum equation - a nonlinear modification of the linear Klein--Gordon equation (that becomes a nonlinear Schr\"odinger equation in  the non relativistic limit). The same quantum prescription results in a nonlinear
Pauli-Schr\"odinger equation when the spin degree of freedom is appropriately included in a non-relativistic treatment. The derivation of a nonlinear Dirac equation to describe a Fluidon corresponding to a fluid composed of relativisitc spin half particles is underway.

The nonlinear quantum mechanics, suggested in this  paper, is likely to be a new and promising tool for investigating fluids whose
dynamics is controlled by  both their quantum and  thermal  aspects. It will probably be the principal step in  developing a basic theory for
matter with extremely high energy-density.

Acknowledgement: Illuminating discussions with Profs. George Sudarshan and Cecile Dewitt are gratefully acknowledged.

\end{document}